\documentclass[referee]{aa}
\usepackage{epsfig}
\usepackage{natbib}
\bibpunct{(}{)}{;}{}{,}

\newif\ifAMStwofonts
\def\spose#1{\hbox to 0pt{#1\hss}}
\def\simlt{\mathrel{\spose{\lower 3pt\hbox{$\mathchar"218$}}
     \raise 2.0pt\hbox{$\mathchar"13C$}}}
\def\simgt{\mathrel{\spose{\lower 3pt\hbox{$\mathchar"218$}}
     \raise 2.0pt\hbox{$\mathchar"13E$}}}

\begin{document}
%\thesaurus{02.08.1; 09.01.1; 11.05.2}
\title{Multiple starbursts in Blue Compact Galaxies}
\author{Simone Recchi\inst{1}\thanks{recchi@ts.astro.it} 
        \and Francesca Matteucci\inst{1}\thanks{matteucci@ts.astro.it} 
        \and Annibale D'Ercole\inst{2}\thanks{annibale@bo.astro.it} 
        \and Monica Tosi\inst{2}\thanks{tosi@bo.astro.it}}
\offprints{F. Matteucci}
\institute{
Dipartimento di Astronomia, Universit\`a di Trieste, Via G.B. Tiepolo, 11,
34131 Trieste, Italy \and
Osservatorio Astronomico di Bologna, Via Ranzani 1, 40127 Bologna, Italy}
\date{Received  /  Accepted   }
\abstract{ 
In this paper we present some results concerning the effects
of two instantaneous starbursts, separated by a quiescent period, on
the dynamical and chemical evolution of blue compact dwarf galaxies.
In particular, we compare the model results
to the galaxy IZw18, which is a very
metal-poor, gas-rich dwarf galaxy, possibly experiencing its first or second
burst 
of star formation.  We follow the evolution of a
first weak burst of star formation followed by
a second more intense one occurring after several hundreds million years. 
We find that a galactic wind develops only during the second burst and that 
metals produced in the burst are preferentially lost relative to the 
hydrogen gas. We predict the evolution of several chemical abundances 
(H, He, C, N, O, $\alpha$-elements, Fe) in the gas inside and outside 
the galaxy, by taking into account in detail
the chemical and energetical contributions from type II and Ia supernovae. 
We find that the abundances predicted for the star forming region are in 
good agreement with the HII region abundances derived for IZw18.
We also predict the abundances of C, N and O expected for the HI gas 
to be compared with future FUSE abundance determinations.
We conclude that IZw18 must have experienced two bursts of star formation, 
one 
occurred $\sim 300$ Myr ago and a present one with an age between 4-7 Myr.
However, by taking into account also other independent estimates, such as 
the color-magnitude diagram and the spectral energy distribution of stars 
in IZw18, and the fact that real starbursts are not instantaneous, 
we suggest that it is more likely that the burst age is between 
4 and 15 Myr.
 }
\maketitle
\bigskip\bigskip\bigskip\bigskip

\section{Introduction}
Blue Compact Dwarf galaxies (BCD) are characterized by compact
appearance, high gas content, very blue colors and low chemical
abundances. These properties are typical of unevolved systems, thus
suggesting that BCD should have suffered very few bursts of star formation
during their lives and that some of them are probably experiencing 
their first burst
(Searle \& Sargent 1972). In a recent
review, Kunth \& \"Ostlin (2000) argued that, despite a few remaining
young galaxy candidates (like IZw18, SBS0335-052 or HS0822+3542;
Lipovetsky et al. 1999; Kniazev et al. 2000), in most BCDs an old
underlying stellar population does exist, revealing at least another
burst of star formation (SF) besides the present one. In a recent
survey of BCD, \"Ostlin et al. (2001) have found that, although the
young burst population dominates the integrated optical luminosities,
it contributes only marginally to the total stellar mass. It is
possible that each starburst episode in BCD be followed by a
quiescent (or almost quiescent) period, with a time scale of the order
10$^8$ to 10$^9$ yr (Leitherer 2001), during which winds produced by
supernova explosions expel the gas out of the region and the star formation
fades. When the luminosity of the burst is no more able to sustain the
wind, the gas should cool and collapse back toward the center of the
galaxy, thus allowing for the onset a new burst (Babul \& Rees 1992;
D'Ercole \& Brighenti 1999). Off-centered SN explosions can also drive
inward-propagating shocks, thus creating the conditions of a new SF
event in the center of the galaxy (Mori et al. 2001).

The galaxy IZw18 is the most metal-poor local galaxy known so far and was
considered until recently as the best candidate for a truly ``young''
galaxy. Stellar population analyses by Hunter \& Thronson (1995) and
Dufour et al. (1996) were not deep enough to reveal any old stellar
population, but recent studies of deeper Color-Magnitude Diagrams
(CMD), both in the optical (Aloisi, Tosi \& Greggio 1999; hereafter
ATG) and in the infrared (\"Ostlin 2000) revealed the presence of two
stellar populations in IZw18: a young population with an age of 
$\sim 15$ Myr and an asymptotic giant branch population with an age of
several $10^{8}$ years.  Chemical evolution models by Kunth et
al. (1995) fit the abundances observed in IZw18 with one, or at
maximum two short bursts. Evolutionary population synthesis models by
Mas-Hesse \& Kunth (1999; hereafter MHK) showed that the present
burst is very young (ranging between 3 and 13 Myr, depending on whether
the burst is instantaneous or continuous) and the contribution of the
stars of an ancient burst to the emission over the whole UV-optical
range, if any, is negligible.  Legrand (2000) and Legrand et
al. (2000) proposed instead a low and continuous SF regime for IZw18;
furthermore they assumed that the observed metals cannot result from
the material ejected by the aging starburst, because these metals are
hidden in a hot phase and therefore undetectable when using the
optical spectroscopy.

In a previous paper (Recchi, Matteucci \& D'Ercole 2001; hereafter
Paper I) we presented a study concerning the effect of a single,
instantaneous starburst on the dynamical and chemical evolution of a
gas-rich dwarf galaxy, with galactic parameters resembling those of
IZw18. We showed that the observed abundances of IZw18 could be
reproduced also by a single-burst model provided that its 
age is $\sim$ 30 Myr,
occurring in a primordial gas (zero initial metallicity), although we
did not exclude the presence of an underlying older population which
polluted only slightly the ISM (less than 1/100 of solar metallicity).
However, the estimate of $\sim 30$ Myr for IZw18 seems to be in contrast 
both with the spectral energy distribution and the color-magnitude diagram, 
as mentioned above.

Other results of Paper I can be summarized as follows:

\par i) a galactic wind develops as a consequence of the starburst and
carries away mostly the metals produced during the starburst. In
particular, we found that the metals produced by type Ia SNe are lost
even more efficiently than the metals produced by type II SNe. This
fact is important since different SN types produce different elements,
in particular, the net effect is to enhance the $\alpha$-element over
Fe ratio inside the galaxy relative to the gas lost from the galaxy.  

\par ii) The cooling of metals in the gas was found to be very
efficient so that most of the metals should be found
in the cold gas phase already
after few Myr from the beginning of the burst.  Both results i) and
ii) depend on the assumed energy transferred from SNe into the
interstellar medium (ISM), a crucial parameter in galaxy evolution studies.

\par In this paper we intend to test the hypothesis of multiple
starbursts with our model. In order to investigate this hypothesis, we
performed some numerical simulations to study in detail the dynamical
and chemical evolution of a gas-rich dwarf galaxy experiencing two
single, instantaneous starbursts. Our approach makes use of a
two-dimensional dynamical model coupled with detailed chemical
enrichment from both type II and type Ia SNe.

In section 2 we summarize the properties of IZw18, in section 3 we
describe the model and the assumptions adopted in our simulation. The
results are presented and discussed in section 4. Finally in section 5
some conclusions are drawn.

\section{IZw18}

The blue compact dwarf galaxy IZw18 (also known as Mrk 116) is the
most metal-poor galaxy known locally (the metallicity is $\sim$ 1/50
Z$_{\odot}$). More metal-deficient BCD have not been found in spite
of extensive surveys (Masegosa et al. 1994; Terlevich et
al. 1996). The colors of this galaxy are exceptionally blue; the most
recent estimates give $U-B=-0.88$ and $B-V=-0.03$ (van Zee et
al. 1998), thus suggesting the presence of a dominant very young
population (van Zee 2001). Given the low degree of pollution, IZw18 is
also an ideal laboratory to measure the primordial helium abundance
(Olive et al. 1997; Skillman et al. 1998; Izotov et al. 1999).

The morphology of IZw18 is rather complex, consisting of a
``peanut-shaped'' main body, formed by two starbursting regions,
associated with two H\,{\sc ii} regions (Dufour et al. 1996). To the
north and west of the main body, there are several diffuse features
noted from the deep ground-based imagery of Davidson et al. (1989) and
Dufour \& Hester (1990), but only the nearest of these objects
(component C in the nomenclature of Davidson et al.  1989, located
22'' NW of the NW region) is a blue irregular galaxy physically
associated with the main body, while the others are background
galaxies (Dufour et al. 1996). For simplicity, we focus our attention
only on the main body features (see Paper I for more references).

The dynamical mass and the H\,{\sc i} mass of the main body of IZw18
are of the order of $\sim 10^{8}$ M$_\odot$ and $\sim  
10{^7}$ M$_\odot$,
respectively (see e.g. Lequeux \& Viallefond 1980; Davidson \& Kinman
1985; Viallefond et al. 1987; Petrosian et al. 1997; van Zee et
al. 1998). The H\,{\sc i} column density is as high as N$_{\ion{H}{i}}
\simeq 2 \times 10^{21}$ cm$^{-2}$ (Vidal-Madjar et al. 2000).

Abundances derived from H\,{\sc ii} regions are reasonably well known
(Searle \& Sargent 1972; Dufour et al. 1988; Skillman \& Kennicutt
1993; Garnett et al. 1995; Garnett et al. 1997; Izotov \& Thuan 1999;
Izotov et al. 1999; Izotov et al. 2001a), and indicate a low content of
oxygen but a rather high N/O ratio. However, there is still some
debate in the literature about the H\,{\sc i} abundances in IZw18. The
first measurement of this kind, by Kunth et al. (1994), indicated a
metallicity which was a factor of 20 lower than in the H\,{\sc ii}
region, although the UV absorption lines they used were saturated, so
this result remains still uncertain (Pettini \& Lipman 1995). Results
by Pettini \& Lipman (1995) and van Zee et al. (1998) seem to indicate
indeed a metallicity comparable to the abundance in the H\,{\sc ii}
regions. A recent work of Levshakov et al. (2001) about argon and
silicon emission line profiles in IZw18 also reveals a good degree of
mixing between hot and cold regions. Future FUSE abundance determinations 
will certainly shed more light on the problem.

\section{The model}

The dark matter content, the interstellar medium (ISM)
initial mass and the mass of stars formed during the first burst 
are shown in Table 1, and are
chosen to match the observed values of IZw18.
We assume that the dark matter has a quasi-isothermal distribution with
a flat profile in the center (see Paper I). 
In fact, Borriello \& Salucci (2001) have recently shown that
rotation curves in dwarf galaxies, as well as in spiral galaxies, are
consistent with a flat central dark matter distribution.

We assume that the first burst of SF produces 10$^5$ M$_\odot$ of
stars, distributed according to a Salpeter IMF, with an upper mass
limit of 40 M$_\odot$ and a lower mass limit of 0.1 M$_\odot$. We
follow the dynamical and chemical evolution of this first burst of SF
up to 300 Myr (model M300) and up to 500 Myr (model M500), solving the
set of gasdynamical equations in two dimensions described in Paper
I. According to the results of ATG, who suggested that the stars
of the present burst could have formed with a flat IMF,
we consider
also an IMF with a slope x$=0.5$ (model M300F). 
The model parameters are shown in Table 2.

The nucleosynthesis prescriptions are summarized in Table
3.The initial metallicity of the gas is set to zero.
The adopted set of yields is Woosley \& Weaver (1995; hereafter WW)
case B and metallicity $1/100 Z_{\odot}$ for massive stars, 
whereas for low and intermediate-mass stars we
consider both the results of Renzini \& Voli (1981; hereafter RV81),
and the more recent set of yields by van den Hoek \&
Groenewegen (1997; hereafter VG97), which has been
successfully tested by chemical evolution models of the Galaxy (Chang
et al. 1999; Romano et al. 2000). One of the main differences between
the two sets of yields is the assumption about the mass loss scaling
parameter $\eta_{\rm AGB}$ (Reimers 1975): in RV81 most of the models
are calculated assuming $\eta_{\rm AGB}=1/3$, while VG97 consider higher
mass loss efficiencies, resulting in lower CNO yields. The adopted mixing 
length
parameter for low and intermediate-mass stars $\alpha_{\rm ML}$ is 
different from
zero, in order to allow the production of primary nitrogen
in intermediate-mass stars.
Concerning type Ia SNe, we adopt the yields of Nomoto, Thielemann
\& Yokoi (1984; hereafter NTY).

The initial abundances of the second stellar generation are simply the
abundances of the cold gas in the central region (a sphere of $\sim$
200 pc of radius) at the onset of the second burst (see Table 2),
which result from the pollution of the first burst. 
We follow the evolution, in space and time, of several chemical elements
of particular astrophysical interest (namely H, He, C, N, O, Mg, Si,
Fe). The supernova (Ia and II) rates and the energetic prescriptions
(explosion energy, thermalization efficiencies) are those adopted
in Paper I, namely $\eta_{\rm II}=0.03$ and $\eta_{\rm Ia}=1$, where
with $\eta_{\rm II}$ and $\eta_{\rm Ia}$ we indicate the efficiency of
energy transfer from SNe into the ISM. The reason of this choice is
that SNeII explode in a cold and dense medium, thus radiating away most
of their internal blast wave energy (cooling is proportional to the
square of the density), whereas type Ia SNe occur after a minimum time delay
of ($\sim$ 30 Myr) in an already hot and diluted medium, thus
transferring most of their energy into the ISM.

\begin{table}
\begin{centering}
\caption[]{First burst of star formation}
\begin{tabular}{cccc}
\noalign{\smallskip}
\hline
\noalign{\smallskip}
M$_*^{I}$ (M$_{\odot})$ & M$_{\rm g}$ (M$_{\odot})$ & 
M$_{\rm h}$ (M$_{\odot})$ &
Z$_{\rm i}$ (Z$_{\odot}$)\\
\noalign{\smallskip}
\hline\noalign{\smallskip}
10$^5$ & 1.7 $\times 10^7$ & 6.5 $\times 10^8$ & 0\\
\noalign{\smallskip}
\hline
\end{tabular}
\vspace{0.3cm}

Notes: M$_{*}^{I}$ is the mass of stars formed in the first episode of
SF, M$_{\rm g}$ is the mass of gas in the galactic region,
M$_{\rm h}$ is the total mass of the dark halo and Z$_{\rm i}$ is the
initial metallicity.
\end{centering}
\end{table} 

\begin{figure}
\centering
\vspace{-0.2cm}
\epsfig{file=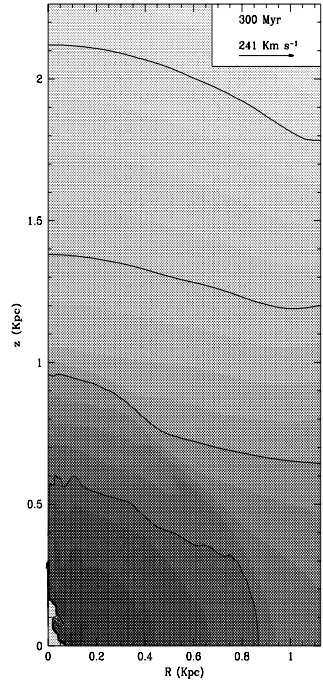, height=10.5cm,width=9cm}
\caption[]{\label{fig:fig 1} Initial gas density profiles of model
M300. The density scale is logarithmic and varies linearly between -29
and -23.5.}
\end{figure}

\begin{figure}
\centering
\vspace{-0.2cm}
\epsfig{file=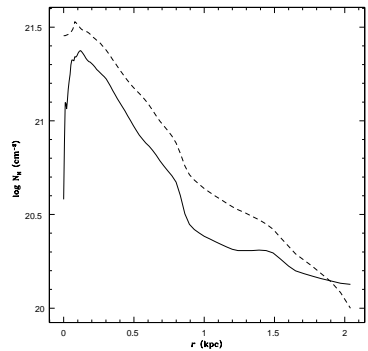, height=5cm,width=5cm}
\caption[]{\label{fig:fig 2} Column density of the initial ISM seen
edge-on (dashed line) and face-on (solid line).}
\end{figure}

\begin{table*}
\begin{centering}
\caption[]{Model parameters}
\begin{tabular}{lcccccc}
\noalign{\smallskip}
\hline
\noalign{\smallskip}
Model & Onset II burst (Myr) & IMF slope & $Age_{burst}$ (Myr) & 
log N$_{\rm H}$ (cm$^{-2}$) &
M$_*^{II}$ (M$_{\odot})$ & Z$_{\rm i}$ (Z$_{\odot}$)\\
\noalign{\smallskip}
\hline\noalign{\smallskip}
M300 & 300 & -2.35 & 159 & -21.5 & 5.8 $\times 10^5$ & 1/50\\
M300F & 300 & -1.5 & 171 & -21.8 & 4 $\times 10^5$ & 1/40\\
M500 & 500 & -2.35 & 154 & -21.4 & 6.25 $\times 10^5$ & 1/100\\
\noalign{\smallskip}
\hline
\end{tabular}
\vspace{0.3cm}

Notes: $\Delta$ t is the duration of the second burst; N$_{\rm H}$ is
the peak of column density in the central region at the onset of the
second burst, M$_{*}^{II}$ is the mass of stars formed in this episode
of SF and Z$_{\rm i}$ is the mean metallicity in this
region.
\end{centering}
\end{table*} 

\begin{table*}
\begin{centering}
\caption[]{Chemical prescriptions}
\begin{tabular}{cccccc}
\noalign{\smallskip}
\hline
\noalign{\smallskip}
Case & SNII yields & IMS yields & SNIa yields & $\alpha_{\rm ML}$ &
$\eta_{\rm AGB}$ \\
\noalign{\smallskip}
\hline\noalign{\smallskip}
R & WW (Z = 1/100 Z$_{\odot}$) & RV81 & NTY & 1.5 & 0.333\\
V & WW (Z = 1/100 Z$_{\odot}$) & VG97 & NTY & 2.0 & 4\\
\noalign{\smallskip}
\hline
\end{tabular}
\vspace{0.3cm}

References: WW Woosley \& Weaver 1995 (models with Z=0.01 Z$_{\odot}$);
RV81 Renzini \& Voli 1981; VG97 van den Hoek \& Groenewegen 1997; 
NTY Nomoto, Thielemann \& Yokoi 1984
\end{centering}
\end{table*}

Before the onset of the second burst we checked if the physical conditions 
of the gas allowed new star formation to set in.
In particular, we controlled whether cold gas was present in the central 
region and whether its 
column density was larger than the threshold value for SF
($\sim$10$^{21}$ cm$^{-2}$), as suggested by  Skillman et al. (1988)
and Sait\=o et
al. (1992). We ran the simulations for $\sim$ 150 Myr from the
beginning of the second burst for all the models M300, M300F and M500.

\section{Results}
\subsection{Dynamical results}
\subsubsection{Model M300}

Owing to the low luminosity of the first burst ($\sim\,3\times \,
10^{37}$ erg s$^{-1}$ during the SNII phase, a factor $\sim$ 10 lower
during the SNIa phase), after 300 Myr the galactic wind still does not
develop. A cold, dense shell with dimensions $\sim$ 200$\times$100 pc
forms and outside this region the ISM is practically unperturbed (see
Fig. 1).  Inside the central region, after 300 Myr are present
5.8$\times 10^6$
M$_\odot$ of cold gas ($< 2 \cdot 10^{4}$K). The mean
metallicity inside this region is $\sim 1/50 Z_{\odot}$,
whereas in the whole galaxy (roughly an ellipsoid with dimensions
700$\times$1000 pc) the gas is more diluted (the mean abundance is
$\sim$ 1/500 Z$_\odot$). Most of the metals in the star formation
region are cold (more than 99 \% of metals have T$<$ 2$\times 10^4$
K). This is because the superbubble evolution is slow and no shell
rupture occurs.  For this reason the internal cavity can cool
significatively (see also Paper I).

The column density in the central region largely exceeds the 
threshold density
(see Figure 2).  Therefore a new SF event can
develop with a pre-enrichment of 1/50 Z$_\odot$. Assuming an
efficiency of SF of 0.1, we obtain for this second burst $\sim$
5.8$\times 10^5$ M$_\odot$ of stars, in agreement with the mass
of stars deduced for IZw18 by assuming an episode of
continuous star formation lasting 13 Myr (MHK).

We assume, in this second burst, a thermalization efficiency slightly
higher than that adopted for the first burst (and in Paper I): in
particular we assume $\eta_{\rm II}=0.05$. In fact, the very central
regions are hot and rarefied, owing to the activity of the first
generation of stars, thus a higher value of this efficiency should be
expected.  We should however consider that star formation is more likely
to be found
in the cold and dense shell, rather than in the central region where
the gas is depleted. The physical conditions of the shell
suggest indeed a value of the thermalization efficiency around 0.05, 
computed as described in 
Bradamante et al. (1998).

In Fig. 3 the dynamical evolution of this model is shown. Because of
the thermodynamical conditions of the central region and the value of
the thermalization efficiency, the impact of the energetic input of
the second generation of stars on the ISM dynamics is rather strong.
In fact, already after 30 Myr a breakout occurs and the
gas produced by this second generation of stars is easily channelled
along the galactic chimney.

The most relevant dynamical results of Paper I indicated a selective loss
of metals, in the sense that the metals produced during the burst were
ejected more easily than the gas originally present, and a fast cooling of
metals, due to the slow evolution of the superbubble, was present. 
In this model
the selective loss of metals is confirmed: at the end of the
simulation ($\sim$ 150 Myr after the onset of the second burst) $\sim$
75 \% of pristine gas (not processed through stars) is lost, 
while the galaxy looses $\sim$ 87\% of
the metals produced in the first generation of stars and $\sim$ 92 \%
of those produced in the second generation. 

Differences between ejection efficiencies of SNIa and SNII products
are still present: SNIa products are more easily channelled along the
galactic funnel, but this effect is less evident compared to the
single- burst model. In fact, the SNII products in this case are lost 
more easily than in the one-burst case. 
This is partly due to the fact that we assume $\eta_{II}$ larger than in 
the one-burst case and partly by the fact that a cavity is already 
carved by the first burst when the second burst sets in. Both these effects
favor a larger loss of metals from type II SNe. 
As a consequence of this, after $\sim$ 20 Myr the fraction
of cold metals inside the galactic region is around 77\%, while in the
single- burst model, with a slower evolution, this fraction was
around 95\%.

\begin{figure*}
\centering
\vspace{-0.2cm}
\epsfig{file=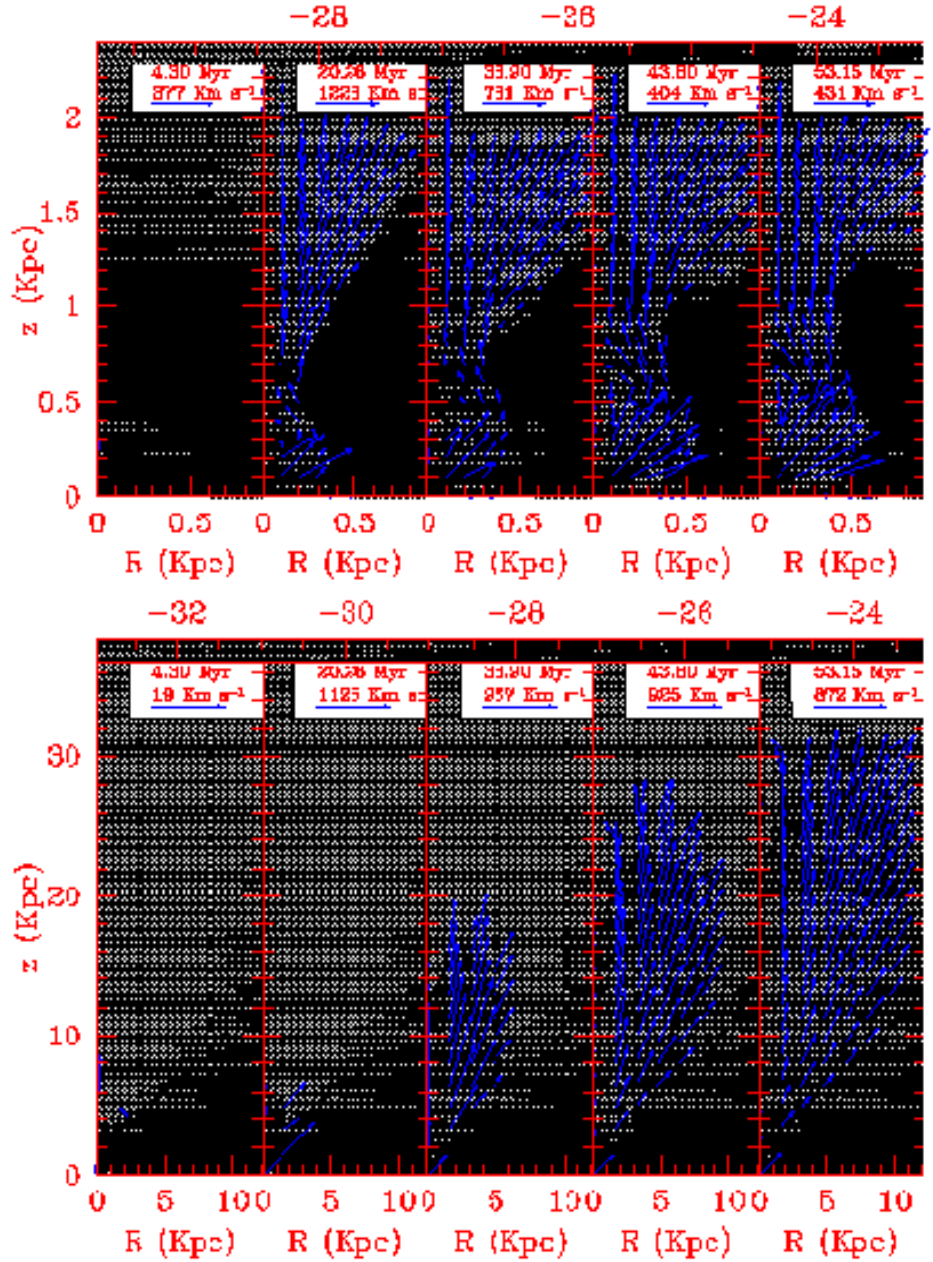, height=24cm, width=18cm}
\vspace{-5cm}
\caption[]{\label{fig:fig 3} Density contours and velocity fields for
model M300 at different epochs (the time labelled on top of each
panel refers to the time elapsed since the beginning of the second
burst). The density scale (logarithmic) is given in the strip on top
of the figures. The lower panels show the whole evolution of the
system, whereas the upper panels are a zoom in the central regions. In
order to avoid confusion, we draw only velocities with values greater
than 1/10 of the maximum value. This is true also for Fig. 4.}
\end{figure*}

\begin{figure*}
\centering
\vspace{-5cm}
\epsfig{file=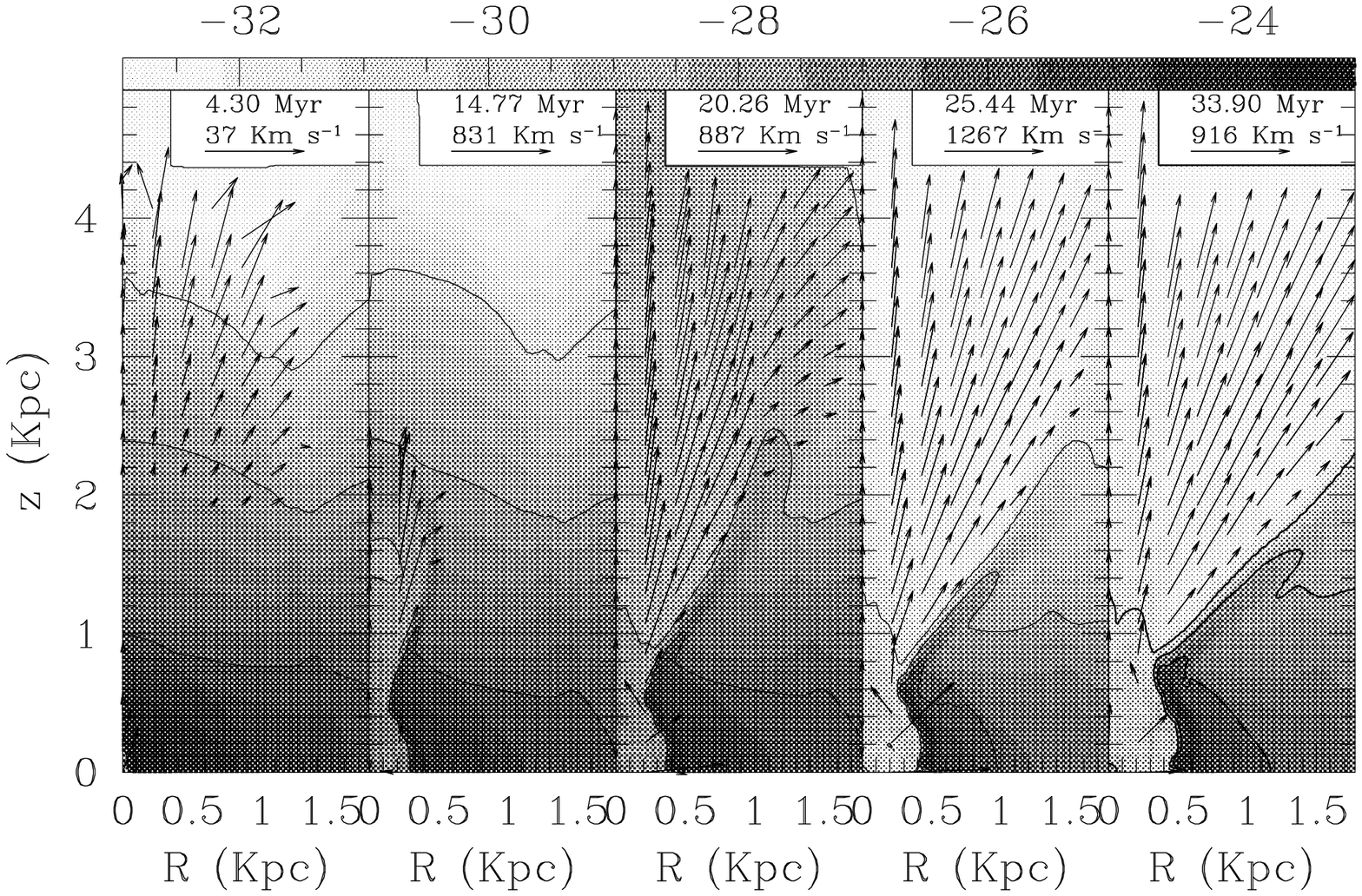, height=15cm,width=15cm}
%\vspace{-5cm}
\caption[]{\label{fig:fig 3} Density contours and velocity fields for
model M300F at different epochs.}
\end{figure*}

\subsubsection{Model M500}

We now follow the hydrodynamical evolution of the first burst of star
formation up to $\sim$ 500 Myr. The evolution in this first phase is
the same as in model M300. The cavity is only slightly narrower and
more elongated along the polar axis; consequently the bubble slightly
deflates and part of the pristine gas outside the bubble moves toward
the center, along the R-direction.  Owing to this, the gas mass in the
star forming region is a little larger than in model M300 ($\sim 6.25
\times 10^6$ M$_\odot$) and the metallicity is reduced (Z $\sim$ 1/100
Z$_\odot$). The mass of stars formed in this second burst is then 6.25
$\times 10^5$ M$_\odot$.

After the onset of the second burst, the evolution of this model
is still similar to that of model M300; a breakout quickly occurs (after
$\sim$ 30 Myr) and the metals produced in this second burst are easily
lost. Approximately 80 \% of metals are in a cold phase after 20
Myr, while if we consider only products of the second generation of
stars, this fraction reduces to 70 \%.  The fraction of unprocessed ISM
ejected in this model is 63.6\%, while the ejection efficiencies 
are $\sim$ 81 \% for the first generation of stars and $\sim$ 91 \% 
for the second generation. 

\subsubsection{Model M300F}

In this model, owing to the flatter IMF, more SNeII are produced. At
the end of the first burst less gas is present in the central region
($\sim 4 \times 10^6\; M_{\odot}$). The stellar population is
strongly biased toward massive stars and $\sim$ 80 \% of the total
stellar mass is in the form of stars more massive than 2 M$_\odot$ (more
than twice of what estimated by MHK). The total number of type II SNe
is around 1.3 $\times\; 10^4$, which is three times the number of
SNeII in model M300.

In Fig. 4 the temporal evolution of model M300F is shown. At the onset
of the second burst (left panel), a weak galactic wind is already
present, at variance with model M300, in which most of the ISM is still
unperturbed (Fig. 1). This is due to the fact that the mean luminosity
of the first burst is three times higher than in model M300 ($\sim$
10$^{38}$ erg s$^{-1}$ during the SNII phase). The impact of the
second burst of SF is thus even stronger than in previous cases 
and already after $\sim$ 15
Myr a breakout occurs.

The luminosity of the second burst is rather high and consequently the
ejection efficiencies, computed at the end of the simulation, are
close to unity. In particular, $81 \%$ of the unprocessed gas, $86.7
\%$ of metal produced by the first generation of stars and $93.9\%$ of
the metals produced by the second stellar generation are ejected out
of the galactic region.

\subsection{Chemical results}

\begin{figure*}
\centering
\vspace{0.4cm}
\epsfig{file=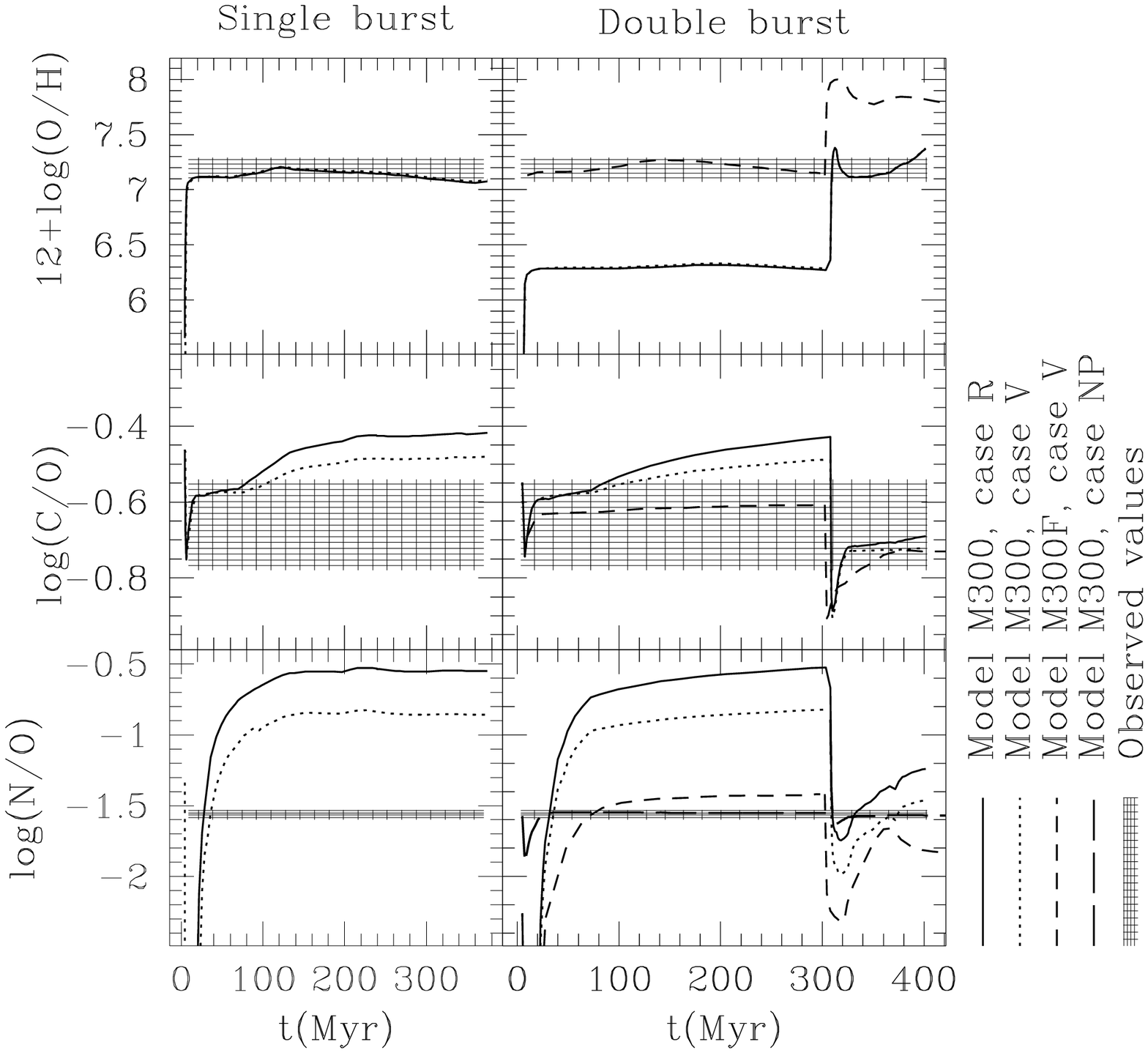, height=15cm,width=15cm}
\vspace{-0.5cm}
\caption[]{\label{fig:fig 5} Evolution of O, C, N for the single-burst
model (left panels) and models M300 and M300F (right panels). In
the bottom-right panel we show the evolution of N/O for a model (model
NP) in which we assumed an ``ad hoc'' production of primary N in
massive stars. The corresponding line is barely visible, because 
is completely embedded  in the observational strip.
The superimposed dashed areas represent the
observative values found in literature for IZw18 (see text for
references).  Left panels represent the evolution of the single-burst
model, with yields by RV81 (solid line) and VG97 (dotted line). In the
right panel the solid line represent the evolution of model M300 with
RV81 yields (case R), the dotted line the VG97 yields (case V), while
the dashed line is the evolution of model M300F, with nucleosynthesis
prescriptions by VG97 (case V)}
\end{figure*}

The evolution of Carbon, Nitrogen and Oxygen for model M300 (cases R
and V) and for model M300F (case V only), compared to the results of
the single- burst model described in Paper I, are shown in Fig. 5. We
have plotted also the evolution of N/O for a model (model NP), similar
to model M300 (case R), but in which we assume an ``ad hoc'' production of
primary nitrogen from massive stars. 
This possibility seems to be suggested by some
stellar evolution models with rotation (see for example Heger \&
Langer 2000). However, precise yields calculations do not seem to
exist, thus we adopted some ``ad hoc'' prescriptions in order to
reproduce the observed N/O in IZw18.  In particular, we assumed that
nitrogen production in massive stars is 1/1000 of the initial mass of
the star, for the whole range 8 $\div$ 40 M$_\odot$ both in the first
and in the second burst. 

The
abundances shown in this plot refer to the abundances of the cold
($T<2\times 10^4$ K) gas inside the galactic region extending for 1
Kpc in the radial direction and for 700 pc in the vertical
direction. They should be compared with the observed HII region abundances.
The abundances of gas in a hotter phase are, in fact,
virtually undetectable (see Paper I for more details). We also
computed the mean metallicity over a region larger than the galactic
one and covering the whole computational grid (roughly 5 Kpc $\times$
2 Kpc) and found a metallicity which can be attributed to the H\,{\sc
i} gas of $Z_{HI} \simeq 1/100 Z_{\odot}$, although this estimate
is rather uncertain with our model.  For this even colder gas, we computed
also the column densities (showed in Fig. 6), integrating along all
the lines of sight, assuming the galaxy to be edge-on (Martin, 1996).  
We found that the abundances of the metals  are roughly a factor of 2 
lower than those predicted for the HII region.
These values can be
used for future comparison with observations (in particular they could
be directly compared with FUSE/HST observations).

Coming back to  Fig. 5, the superimposed dashed area represents the abundances
observed in IZw18 H\,{\sc ii} regions (Dufour \& Hester (1990);
Skillman \& Kennicutt (1993); Izotov et al. (1997); Garnett et
al. (1997); Vilchez \& Iglesias-P\'aramo (1998); Izotov \& Thuan
(1999)). We find that the single-burst model, occurring in a
primordial gas, at an age of $\sim 31$ Myr (case R) can reasonably
reproduce the abundances measured in IZw18. However, as evident in
Fig. 5, the correct N/O ratio would last only for a very short time
(of the order of a couple of Myr) since for $t \simgt$ 31 Myr the N/O
ratio will start to increase, due to the N produced by intermediate
and low mass stars, outside the permitted observational range.
Moreover, this age estimate is inconsistent with other independent estimates
derived from observations (see section 5).

In model M300, at the onset of the second burst, the metallicity
in the central region is $ \sim Z_{\odot}/50$, and a very good
agreement with the observed abundances is obtained, especially for the
O abundance and the C/O ratio, for a larger range of burst ages. The
yields of VG97 produce substantially less CNO elements, compared to
what is obtained by RV81. Therefore, the evolution of nitrogen is
strongly affected by the choice of nucleosynthetic prescriptions,
whereas the oxygen evolution is almost identical, because O is mostly
produced by Type II SNe (see Fig. 5, upper panels).

For the model M300 we find two solutions
reproducing the observed N/O abundance ratio. The first is around 5-6
Myr after the beginning of the second burst and is almost independent
of the the adopted nucleosynthetic prescriptions. The second solution
predicts a larger burst age and depends on the adopted set of
chemical yields: in the R case, the N/O ratio observed in IZw18 is
reproduced after a time between 20 and 40 Myr from the onset of the
second burst, whereas in case V
we need an age between $\sim$ 45 and $\sim$ 70 Myr to fit
the observed abundance ratios. Unfortunately, both these burst age 
estimates are too large and inconsistent with the spectral 
energy distribution. 
The M300F model with a flatter IMF
produces much more oxygen in the first burst and indicates that
the only possible solution is an extremely short age of the second
burst (around 4 Myr, in agreement with the results of MHK).

These results depend also on the thermalization efficiency, in the
sense that if we assume higher $\eta_{\rm II}$, the galactic wind is
stronger and more metals are lost through the galactic chimney, thus
diminishing the abundances inside the galaxy. In the limiting case in
which $\eta_{\rm II}=1$ we have shown in Paper I that the galaxy is
quickly devoided of gas, at variance with what observed in IZw18. Thus
the thermalization efficiency could be higher than the assumed value,
but not too high to devoid the galaxy of gas.  The thermalization
efficiency of SNe, in particular Type II SNe, is a crucial parameter
and a more refined model will be presented in a future paper (Recchi
et al., in preparation).

It is worth noting that we assumed mainly secondary N production
from massive stars, as implied by the adopted yields of WW, and this
means a slow growth of N with time. However, Izotov \& Thuan (1999) and 
Izotov
et al. (2001b) found that some low metallicity BCD show a constant N/O
ratio as a function of O/H. This result can be interpreted as due to
the fact that N from  massive stars is mainly a primary element, 
as originally
suggested by Matteucci (1986).
We explored this alternative and the
resulting N/O evolution as a function of
time, shown in Fig. 5, is rather flat and 
consistent with the nitrogen abundance
observed in low-metallicity BCD galaxies.
In this case, N/O cannot be used as a clock for the burst age.

The evolution of the C/O ratio at the onset of the second burst is
different compared to the behaviour of the N/O ratio. In particular,
all the models shown in Fig. 5 seem to fall to a constant C/O value
(around $-0.9$) at the beginning of the second burst, then rising with
different slopes, at variance with what happens with the N/O
evolution. The reason for this is that carbon is produced also in
massive stars, thus the C/O ratio after the onset of the second burst
is mostly due to the massive stars produced by the second stellar
generation, whereas the level of N is only determined by the IMS
produced during the first burst.

In Fig. 7 the evolution of carbon, nitrogen and oxygen for model M500,
case R, is shown. For this model we obtain good agreement with
the observed abundances
for an age of the secod burst
of $\sim$6 Myr or between $\sim$ 40 Myr and $\sim$ 120 Myr,
a far too large estimate. 
The results of this model slightly
differ from those of model M300; in particular the evolution of
nitrogen appears to be flatter than in model M300. In fact, at the
onset of the second burst, the first generation of stars still
produces and ejects metals into the ISM and the production of nitrogen
due to the first starburst is larger for model M300. Moreover, the
differences in the initial distribution of the ISM and in the
luminosity of the starburst in models M300 and M500, have different dynamical
consequences, which can explain the different N/O behaviours. In
particular, after $\sim$ 150 Myr the ejection efficiency of N is
$\sim$ 0.91 for the M500 model and $\sim$ 0.85 for the M300, thus most
of N produced in model M500 is lost and the evolution of N/O flattens.

\begin{figure}
\centering
\vspace{0.8cm}
\epsfig{file=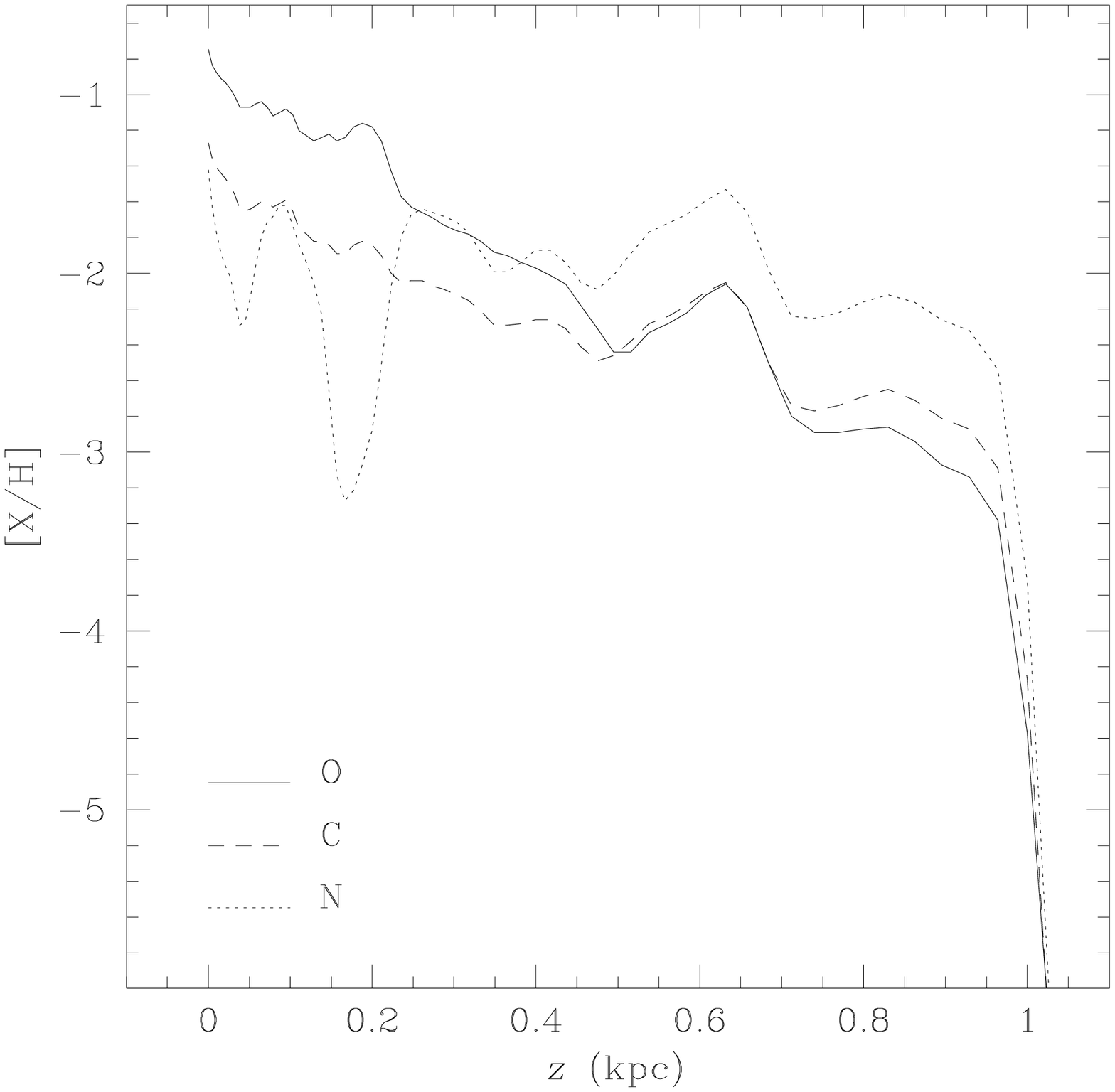, height=4.8cm,width=6cm}
%\vspace{-0.5cm}
\caption[]{\label{fig:fig 6} {Column densities of oxygen (solid line), 
carbon (dashed line) and nitrogen (dotted line) relative to hydrogen, for 
the neutral medium for the model M300 after $\sim$ 6 Myr.  
These values are calculated integrating along all the lines of sight, 
covering the whole computational grid, assuming the galaxy to be edge-on.}}
\end{figure}

\begin{figure}
\centering
\vspace{2cm}
\epsfig{file=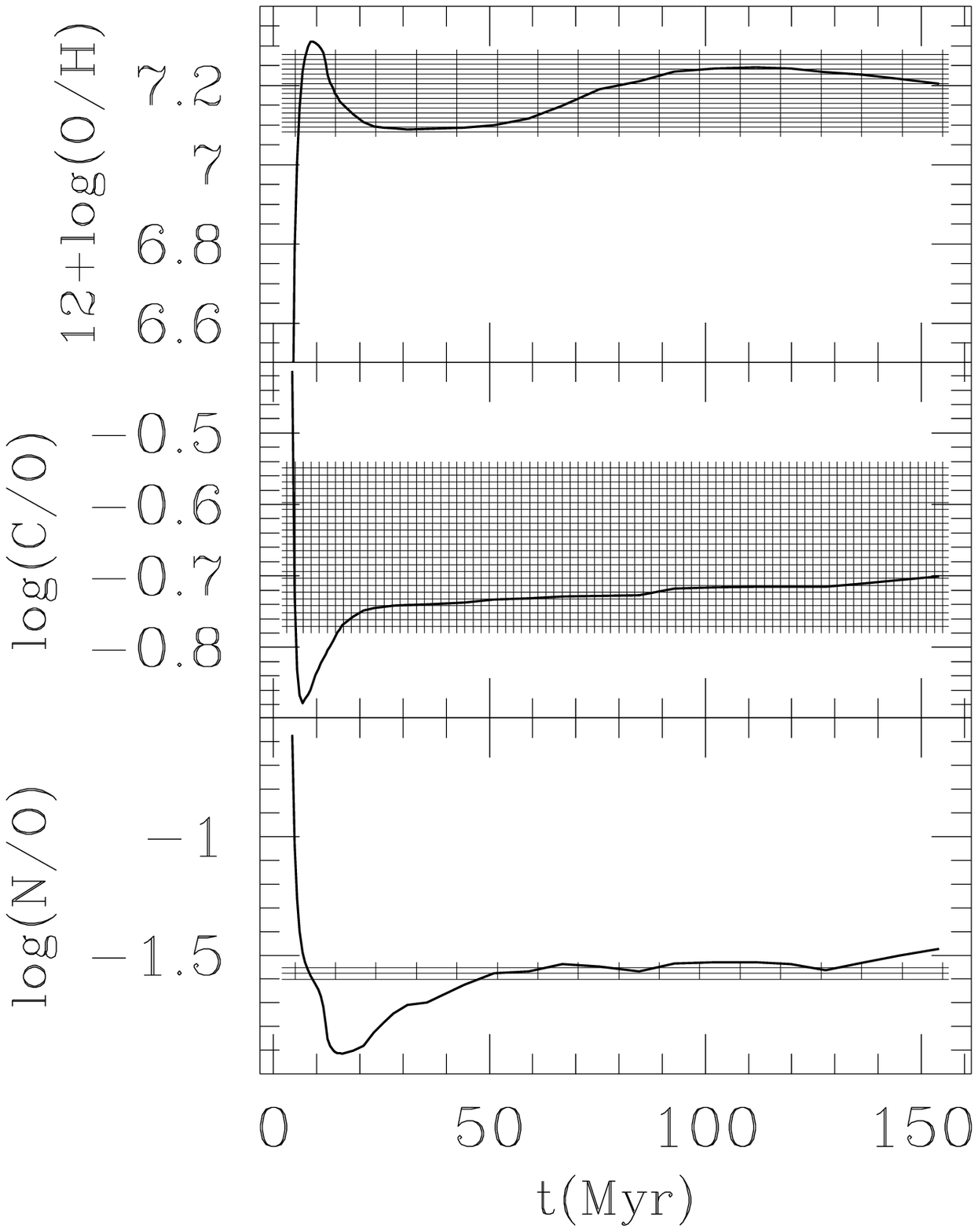, height=7.5cm,width=9cm}
\vspace{-0.5cm}
\caption[]{\label{fig:fig 7} Evolution of O, C, N for model M500. The
time is calculated from the onset of the second burst. As in Fig. 5,
the superimposed dashed areas represent the observative values found
in literature for IZw18}
\end{figure}

In Fig. 8 the [$\alpha$/Fe] ratios for model M300, case R, compared to
the single-burst model is shown. The [O/Fe] ratios in the two-burst case are
always lower outside than inside the galaxy, but the effect is less
evident than in the one-burst case. Note that, after the first burst
in model M300, no outflows are expected, thus no metals are present in
the external region. The evolution of the [O/Fe] ratio in the model
M500 is shown in Fig. 9.

As evident from these figures, the overabundance of $\alpha$-elements
lasts only for the first 20-30 Myr after the second burst and the
maximum overabundance is of the order of 0.4 dex, which is in good
agreement with some observations (Izotov et al. 1997 give [O/Fe] =
0.39 $\pm$ 0.09). Recently, Levshakov et al. (2001) with their
mesoturbulent approach, found higher overabundances of $\alpha$
elements (they found [Si,Ar/Fe] = 0.7 $\pm$ 0.1). Since there are some
indications that WW yields of iron are overestimated by a factor of 2
(Timmes et al. 1995), we tried to calculate $\alpha$/Fe ratios with
reduced iron yields from massive stars and the results for [Mg/Fe] and
[Si/Fe] evolution are shown in Fig. 10.  As shown in this figure,
after the second burst, the overabundances of Si and Mg are larger
than 0.6 dex and they last until $\sim$ 30 Myr after the onset of the
second burst.

As mentioned in the introduction, IZw18 provides useful information
about the primordial helium abundance, because BCD are the least
chemically evolved galaxies known, so they are supposed to contain
very little helium processed by stars after the big bang. Indeed, in
our models, the variations of the helium mass fraction are at maximum
of the order of 1 \%, thus confirming the robustness of this approach
in determining the primordial $^4$He abundance.

\begin{figure}
\centering
\vspace{1.4cm}
\epsfig{file=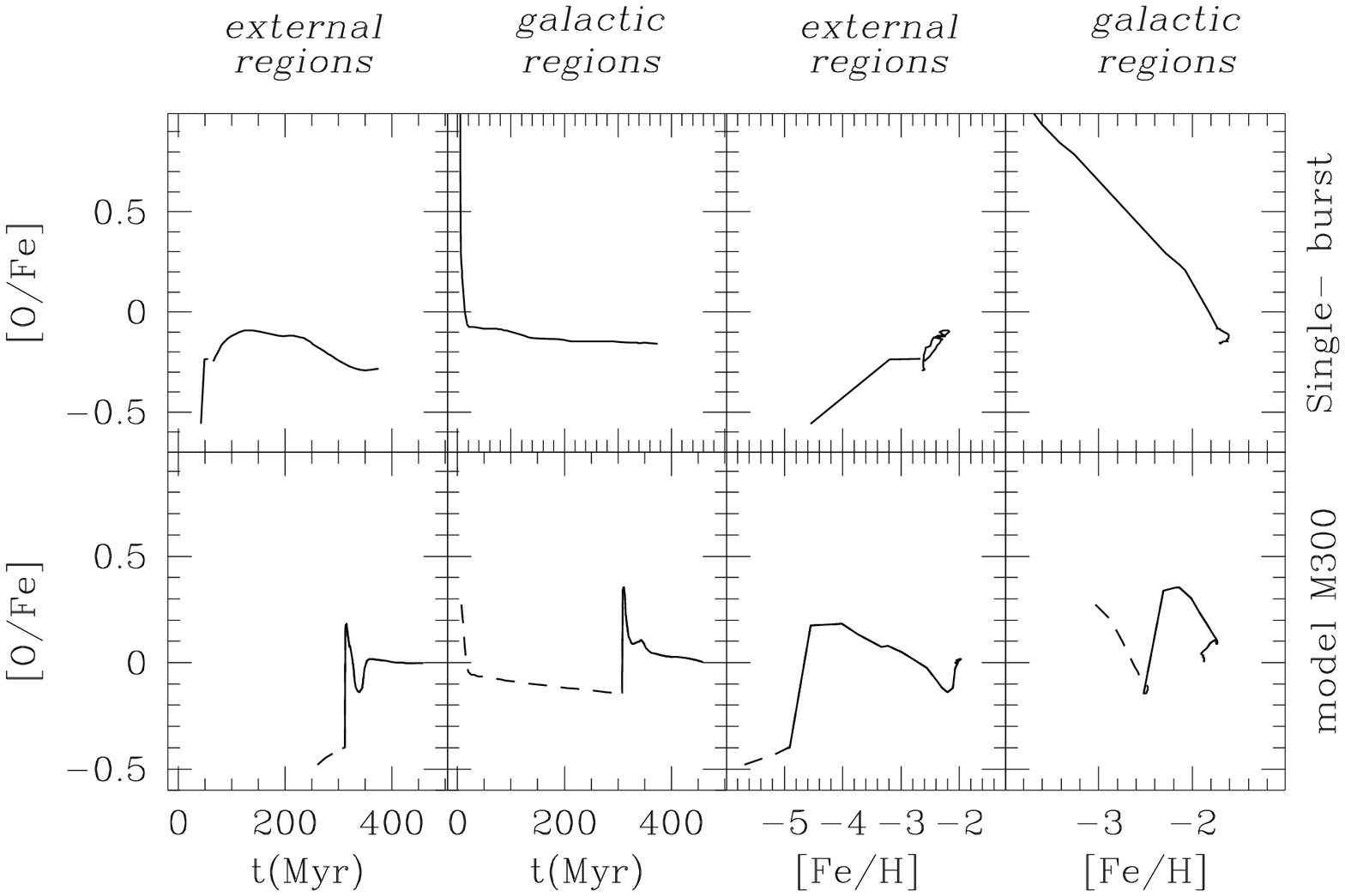, height=7.5cm,width=9cm}
\vspace{-1.5cm}
\caption[]{\label{fig:fig 8} O/Fe vs. time and vs. Fe/H for the
single- burst model (upper panels) and for model M300, case R
(lower panels). Dashed lines represent the chemical evolution of the
first burst for model M300.}
\end{figure}

\begin{figure}
\centering
\vspace{-0.8cm}
\epsfig{file=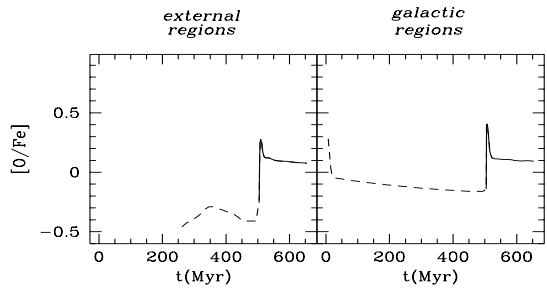, height=7.5cm,width=9cm}
%\vspace{-0.5cm}
\caption[]{\label{fig:fig 9} O/Fe vs. time for the model M500}
\end{figure}

\begin{figure}
\centering
\vspace{0.4cm}
\epsfig{file=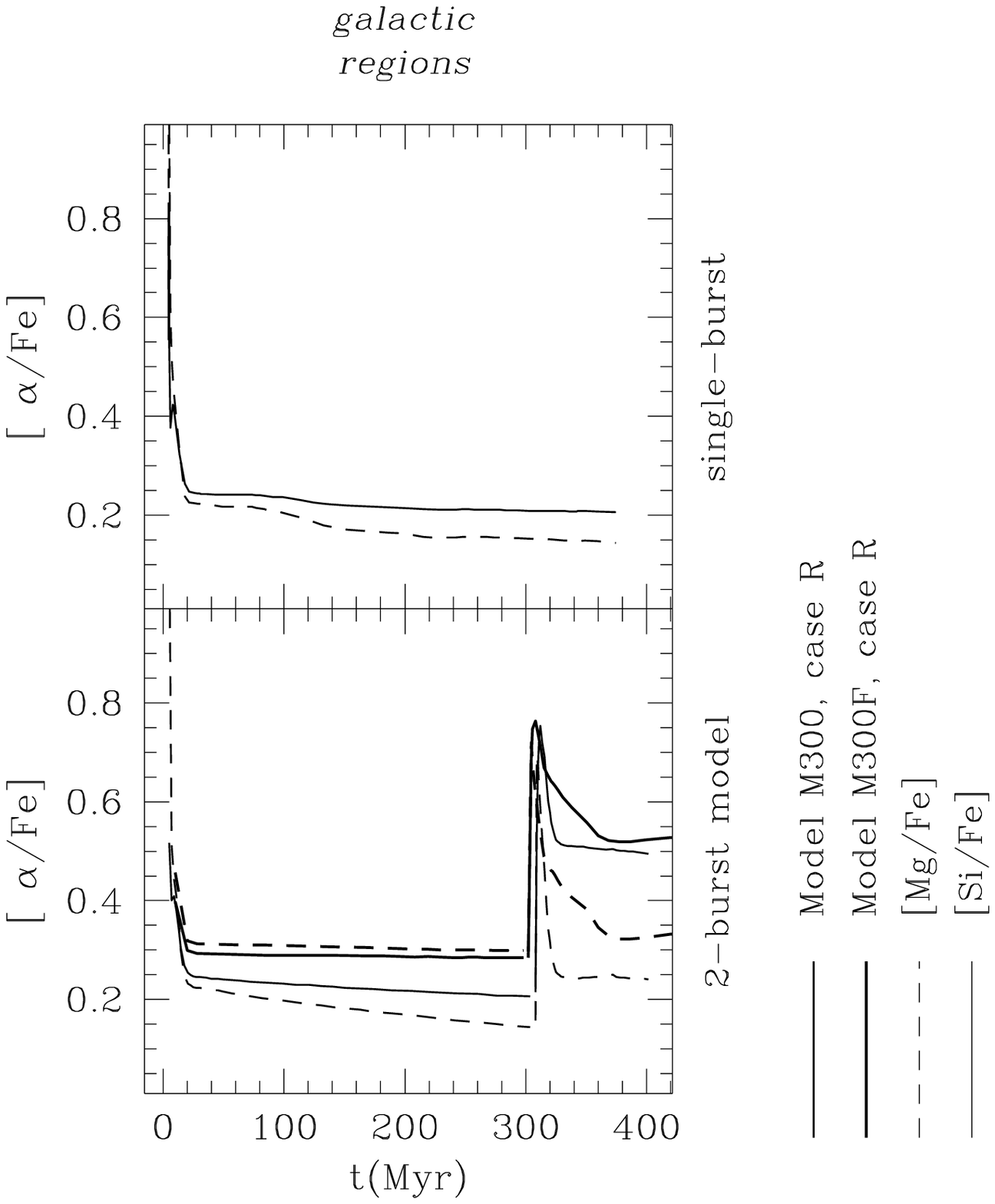, height=7.5cm,width=9cm}
%\vspace{-0.5cm}
\caption[]{\label{fig:fig 10} [Mg/Fe] and [Si/Fe] vs. time for the
single- burst model (upper panel) and for model M300, case R and
M300F, case R (lower panel), calculated with halved Fe yields from
massive stars. Dashed lines represent the chemical evolution of the
magnesium, while solid lines are the evolution of silicon.}
\end{figure}

\section{Discussion and Conclusions}

In this paper we have studied the chemo-dynamical evolution of the gas as 
a consequence of two instantaneous starburts, separated by a quiescent 
period,
in a galaxy similar to IZw18.
We have taken into account in detail both SN (Ia and II) feed-back and 
stellar yields.
The main conclusions 
can be summarized as follows:

\begin{itemize}

\item in most of the explored cases the wind develops only during the
second burst and the newly formed metals are ejected more efficiently
than the pristine gas (mostly H), thus confirming previous results
(Paper I; D'Ercole \& Brighenti 1999; MacLow \& Ferrara 1999).

\item In particular, the metals produced by type Ia SNe are lost more
efficiently than those produced by type II SNe, due to
the higher efficiency in energy transfer by type Ia SNe, which explode
in an already hot and rarefied medium. However this effect, already
found in Paper I where only one burst was assumed, is milder in the
case of 2 bursts, since SNeII in the second burst are also likely to
transfer more energy into the ISM.

\item Our best model suggests that a first weak burst of star formation,
occurred roughly 300 Myr ago, followed by a more intense one, can well
reproduce the properties of IZw18.

\item The predicted chemical abundances for the gas which remains
inside the star-forming region are in very good agreement with 
the HII region
abundances of
IZw18, especially for what concerns the C/O and N/O ratios. The
agreement with the data of IZw18 is definitively better for the case
with 2 bursts than with only one, thus confirming the presence of old
underlying stellar populations (ATG, \"Ostlin 2000). These results are
in good agreement with observations either when using the chemical
yields from low and intermediate mass stars
by Renzini \& Voli (1981) or those by van den Hoek \&
Groenewegen (1997).

\item In the framework of
the standard nucleosynthesis in massive stars we find two solutions for
the age of the second burst: 1) an age of some tens of Myr (depending
on the nucleosynthetic prescriptions adopted) and 2) an age of $4-6$
Myr. This second solution is in good agreement with the spectral
energy distribution models studied by MHK.  
If we assume that massive stars produce only primary N,
the situation changes since the right N/O ratio is achieved already 
after 3-4
Myr after the first burst. Therefore, in this case we cannot use
the N/O as a cosmic clock. However, the assumption of massive stars
producing only primary N is perhaps too extreme.
Therefore, we conclude that on the basis of only chemo-dynamical results we cannot safely 
suggest the age of the present time burst. 
As a consequence of this, we checked our 
burst age by computing the spectral energy distribution expected for the 
second burst, in the best model case. To do this
we have computed $U-B$ and $B-V$ colors by adopting the 
package Starburst99 (Leitherer et al. 1999), a web based software and
data package designed to model spectrophotometric and related
properties of star-forming galaxies.  The results obtained with
Starburst99 were then compared
with the observed values of IZw18 (namely $U-B =
-0.88 \pm 0.06$ and $B-V = -0.03 \pm 0.04$; van Zee et al. 1998) and they are
shown in Fig. 11.  From this figure one can see that
the outputs of Starburst99 are consistent with the
observed $U-B$ only if the second burst has an age between 5.3 and 11
Myr, whereas the $B-V$ color is well reproduced if the age is between
23 and 44 Myr.  However, the $B-V$ color is affected by the presence
of the older stellar population (impossible to simulate with the
Starburst99 package), whereas the $U-B$ color is dominated by young
stars.  Thus,  these results seem to indicate that a very young
age for the second burst (of the order of 5 -- 6 Myr) is the most
likely solution, at least in the hypothesis of an instantaneous burst of
star formation. 
Moreover, we can estimate the age of the second burst also from dynamical 
considerations.
In particular, in our best model the bubble has travelled a distance 
of 720 pc in only 6-7 Myr from the beginning of the starburst.
This distance is exactly the space between the shell and the center of 
the NW HII region observed in IZw18 (Martin, 1996). Martin (1996) herself
suggests an age larger than the one derived here (see Table 4).
This difference is  
due to the  fact that Martin (1996) assumed 
an ISM distribution unmodified by the previous burst.
In fact,  the bubble expansion
timescale strongly depends on the gas distribution.
In conclusion, our results combined with the information we have from 
other studies, as summarized in Table 4, and with the fact that real
starbursts are not instantaneous,
suggest that one can reasonably conclude that the age of the present 
burst should be 
in the range 5-15 Myr.

\item Models with a flat ($x=0.5$) IMF produce much more oxygen after
the first burst of star formation compared to models with the Salpeter
(1955) IMF. For these models the only possible solution is a second
burst of a very short age (around 4 Myr).

\item The [$\alpha$/Fe] ratios in the two burst case are always lower
in the gas lost through the wind than in the gas which remains bound
to the galaxy, thus creating an asymmetry. However this effect,
already present in the one burst case, is here less strong.  
The
[$\alpha$/Fe] inside the galaxy is predicted to be higher than solar
(overabundance of $\alpha$-elements) during the first 30 Myr from the
burst.  In particular, we predict a value of [O/Fe]$\sim$+0.4 dex when
the standard yields are adopted and a value of [O/Fe] $\sim +0.7-0.8$
dex when the Fe in massive stars is artificially lowered by a factor
of two, since the Woosley \& Weaver (1995) yields for Fe seem to be
overestimated.  The last value of [O/Fe] is in very good agreement
with the abundance analysis of IZw18 by Levshakov et al. (2001).

\item Finally, we computed the abundances of metals
which should pertain to the HI gas
and estimated that they are roughly a factor of 2 lower than the abundances 
in the star forming region (HII), a prediction which should be tested on 
new FUSE data. 

\end{itemize}

\begin{table}
\begin{centering}
\caption[]{Ages of the second burst from various sources}
\begin{tabular}{ccc}
\noalign{\smallskip}
\hline
\noalign{\smallskip}
age (Myr) & source & reference \\
\noalign{\smallskip}
\hline\noalign{\smallskip}
5-6 & model M300 & this paper \\
40-70 & model M300 & this paper \\
$\sim$ 4 & model M300F & this paper \\
$\sim$ 6 & model M500 & this paper \\
40-120 & model M500 & this paper \\
6-7 & bubble dynamics (M300) & this paper\\
\hline
5-11 & U-B & S99 \\
23-44 & B-V & S99 \\
13.6 & $M_B$ & S99 \\
\hline
15-20 & optical CMD & ATG \\
15-20 & NIR CMD & Ostlin 2000\\
15-27 & bubble dynamics & Martin 1996 \\
3-13 & integrated spectra & MHK \\

\noalign{\smallskip}
\hline
\end{tabular}
\vspace{0.3cm}

Notes: We quote as S99 our own application of the Starburst99 (Leitherer
et al. 1999) models to our examined cases. See text
for details.
\end{centering}
\end{table}

\begin{figure}
\centering
\vspace{1.4cm}
\epsfig{file=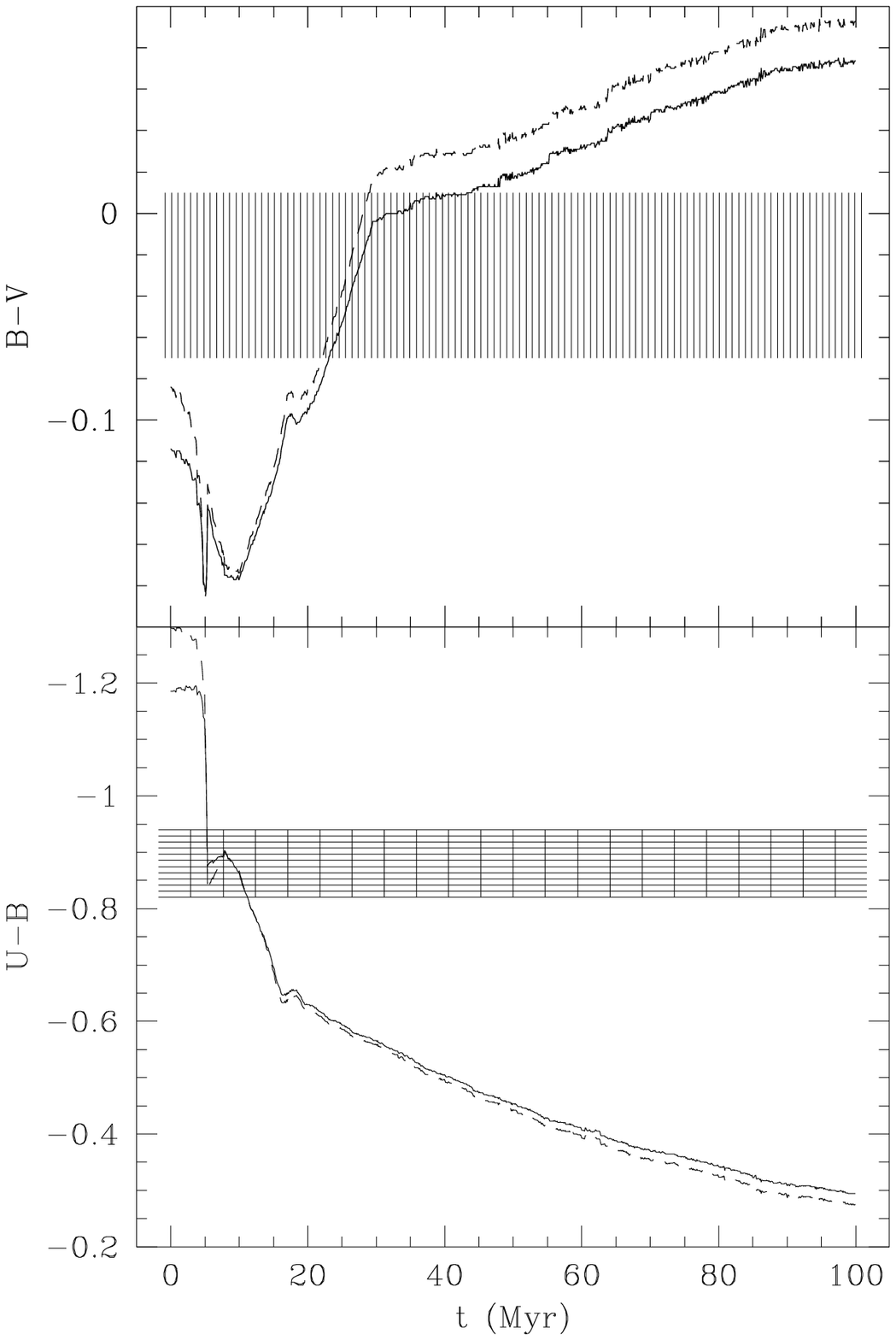, height=7.5cm,width=9cm}
%\vspace{-0.5cm}
\caption[]{\label{fig:fig 11} {$U-B$ and $B-V$ colors computed 
by means of Starburst99 (Leitherer et al. 1999), 
for the model M300 (solid line) and M300F (dashed line), compared 
with published observed values for IZw18, with relative error-bars 
(dashed areas).}}
\end{figure}

Before concluding we want to address some comments about the
limits of the present model.
In particular, the above results refer to the case of two subsequent
instantaneous bursts of SF, the most common scenario usually
attributed to the evolution of BCD.  Until recently BCDs were
supposed to undergo short and intense bursts of star formation,
separated by long quiescent intervals, whereas dwarf irregulars and
giant irregulars appeared to have a more continuous SF activity
(e.g. Tosi 1999). For this reason, the vast majority of chemical
evolution models for BCDs (e.g. Matteucci \& Tosi 1985, Pilyugin 1993,
Bradamante, Matteucci, D'Ercole 1998) were computed assuming burst
durations of 10$^8$ years or less.

Nowadays, however, there is increasing evidence that also BCDs have
rather a {\it gasping} SF, with long episodes of activity, separated
by short quiescent intervals, if any. This continuity has been
hypothesized by a few authors (e.g. Carigi et al. 1995, Legrand 2000),
and has been supported by observations when the Hubble Space Telescope
has allowed to resolve faint single stars even in galaxies outside the
Local Group.  
The application of the method
of synthetic color-magnitude diagrams (CMD) to derive the SF history in 
relatively distant BCD 
has provided fundamental information on the evolution of these
systems.  Deep CMD have been obtained 
from HST optical and infrared
photometry of several BCD by different groups (e.g. Aloisi et
al. 1999, Schulte-Ladbeck et al. 2001, Tosi et al. 2001) and have
provided similar scenarios for the SF histories of the examined
objects: the SF activity has started long ago (in all cases, at least
as long ago as given by the maximum lookback time corresponding to the
depth of the available photometry), but has been very intense only in
a few galaxies. Standard episodes have a duration of several 10$^8$
years and tend to overlap each other with no real quiescent phases in
between, or fairly short ones (lasting only 5--10 Myr).

If the latter scenario turns out to apply in general to most BCD, the
next fundamental step for a better understanding of their chemical
evolution will be to take into account the effect on the ISM of SN
explosions produced during long-lasting SF episodes. 
In fact, in most of the model solutions selected here, 
the predicted N abundances
cross the observed range in a very short time interval (c.f. Figs. 5 and 7).
We expect that models with continuous SF will predict abundances remaining 
within the observed range for longer, more realistic timescales.
Despite the much
heavier computer requirements, we are currently working on this
implementation in our treatment and the results will be presented in a
forthcoming paper.

\begin{acknowledgements}

We like to thank the referee Dr. James Lequeux for his competent and
useful suggestions that helped to significantly improve the clarity of
this paper.  This work has been partly supported by the Italian MIUR
through Cofin2000.

\end{acknowledgements}


\begin{thebibliography}{}
\bibitem[Aloisi, Tosi \& Greggio 1999]{atg}
Aloisi, A., Tosi, M., \& Greggio, L. 1999, AJ, 118, 302 (ATG)
\bibitem[Babul \& Rees 1992]{br}
Babul, A., \& Rees, M.J. 1992, MNRAS, 255, 346
\bibitem[Borriello \& Salucci 2001]{bs_p}
Borriello, A., \& Salucci, P. 2001, MNRAS, 323, 285
\bibitem[Bradamante, Matteucci \& D'Ercole 1998]{b98}
Bradamante, F., Matteucci, F., \& D'Ercole, A. 1998, A\&A, 337, 338
\bibitem[Carigi et al. 1995]{carigi95}
Carigi, L., Colin, P., Peimbert, M., \& Sarmiento, A. 1995, ApJ 445, 98
\bibitem[Chang et al. 2000]{c00}
Chang, R.X., Hou, J.L., Shu, C.G., \& Fu C.Q. 1999, A\&A, 350, 38
\bibitem[Davidson \& Kinman 1985]{dk85}
Davidson, K., \& Kinman, T.D. 1985, ApJS, 58, 321
\bibitem[Davidson, Kinman \& Friedman 1989]{dkf_p}
Davidson, K., Kinman, T.D., \& Friedman, S.D. 1989, AJ, 97, 1591
\bibitem[D'Ercole \& Brighenti 1999]{db}
D'Ercole, A., \& Brighenti, F. 1999, MNRAS, 309, 941
\bibitem[Dufour, Esteban \& Casta\~neda 1996]{d96_p}
Dufour, R.J., Esteban, C., \& Casta\~neda, H.O. 1996, ApJ, 471, L87
\bibitem[Dufour, Garnett \& Shields 1988]{dgs}
Dufour, R.J., Garnett, D.R, \& Shields, G.A. 1988, ApJ, 332, 752
\bibitem[Dufour \& Hester 1990]{dh_p}
Dufour, R.J., \& Hester, J.J. 1990, ApJ, 350, 149
\bibitem[Garnett et al. 1995]{g95}
Garnett, D.R., Dufour, R.J., Peimbert, M., et al. 1995, ApJ, 449, L77
\bibitem[Garnett et al. 1997]{g97}
Garnett, D.R., Skillman, E.D., Dufour, R.J., \& Shields, G.A. 
1997, ApJ, 481, 174
\bibitem[Heger \& Langer 2000]{hl00}
Heger, A., \& Langer, N., 2000, ApJ, 544, 1016
\bibitem[Hunter \& Thronson 1995]{ht}
Hunter, D.A., \& Thronson, H.A. 1995, ApJ, 452, 238
\bibitem[Izotov et al. 2001b]{i01}
Izotov, Y.I., Chaffee, F.H., Foltz, C.B., et al. 2001b, 
Proceedings of the XVII IAP Colloquium "Gaseous Matter in Galaxies and 
Intergalactic Space" ({\tt astro-ph/0109519})
\bibitem[Izotov et al. 1999]{i99b}
Izotov, Y.I., Chaffee, F.H., Foltz, C.B., et al. 1999, ApJ, 527, 757
\bibitem[Izotov, Schaerer \& Charbonnel 2001]{isc}
Izotov, Y.I., Schaerer, D., \& Charbonnel, C. 2001a, ApJ, 549, 878
\bibitem[Izotov \& Thuan 1999]{it99}
Izotov, Y.I., \& Thuan, T.X. 1999, ApJ, 511, 639
\bibitem[Izotov, Thuan \& Lipovetsky 1997]{itl}
Izotov, Y.I., Thuan, T.X., \& Lipovetsky, V.A. 1997, ApJS, 108, 1
\bibitem[Kniazev et al. 2000]{ketal00}
Kniazev, A.Y., Pustilnik, S.A., Masegosa, J., et al. 2000, A\&A, 357, 101
\bibitem[Kunth et al. 1994]{k94_p}
Kunth, D., Lequeux, J., Sargent, W.L.W., \& Viallefond, F. 1994, A\&A, 282, 709
\bibitem[Kunth, Matteucci \& Marconi 1995]{kmm_p}
Kunth, D., Matteucci, F., \& Marconi G. 1995, A\&A, 297, 634
\bibitem[Kunth \& \"Ostlin 2000]{ko_p}
Kunth, D., \& \"Ostlin, G. 2000, A\&AR, 10, 1
\bibitem[Legrand 2000]{l00_p}
Legrand, F. 2000, A\&A, 354, 504
\bibitem[Legrand et al. 2000]{letal_p}
Legrand, F., Kunth, D., Roy, J.-R., Mas-Hesse, J.M., \& Walsh, J.R. 2000, 
A\&A, 355, 891
\bibitem[Leitherer 2001]{l2001}
Leitherer, C. 2001, to appear in the proceedings of ``Astrophysical 
Ages and Time Scales'', ASP Conference Series (eds. T. von 
Hippel, N. Manset \& C. Simpson)
\bibitem[Leitherer et al. 1999]{l99}
Leitherer, C., Schaerer, D., Goldader, J.D., et al., 1999, ApJS, 123, 3
\bibitem[Lequeux \& Viallefond 1980]{lv80}
Lequeux, J., \& Viallefond, F. 1980, A\&A, 91, 269
\bibitem[Levshakov, Kegel \& Agafanova 2000]{lka_p}
Levshakov, S.A., Kegel, W.H., \& Agafanova, I.I. 2001, A\&A, 373, 836
\bibitem[Lipovetsky et al. 1999]{lipo}
Lipovetsky, V.A., Chaffee, F.H., Izotov, Y.I., et al. 1999, ApJ, 519, 177
\bibitem[MacLow \& Ferrara 1999]{mf99}
MacLow, M.-M., \& Ferrara, A. 1999, ApJ, 513, 142
\bibitem[Masegosa, Moles \& Campos-Aguilar 1994]{mmc}
Masegosa, J., Moles, M., \& Campos-Aguilar, A. 1994, 420, 576
\bibitem[Mas-Hesse \& Kunth 1999]{mk_p}
Mas-Hesse, J.M., \& Kunth, D. 1999, A\&A, 349, 765 (MHK)
\bibitem[Matteucci 1986]{m86}
Matteucci, F. 1986, PASP, 98, 973
\bibitem[Matteucci 1996]{m96}
Matteucci, F. 1996, Fund. Cosm. Phys., 17, 283
\bibitem[Matteucci \& Tosi 1985]{mt85}
Matteucci, F. \& Tosi, M. 1985, MNRAS 217, 391
\bibitem[Mori, Ferrara \& Madau 2001]{mfm}
Mori, M., Ferrara, A., \& Madau, P. 2001, {\tt astro-ph/0106107}
%\bibitem[Navarro, Frenk \& White 1996]{nfw_p}
%Navarro, J.F., Frenk, C.S., \& White, S.D.M. 1996, ApJ, 462, 563
\bibitem[Nomoto, Thielemann \& Yokoi 1984]{nty_p}
Nomoto, K., Thielemann, F.-K., \& Yokoi, K. 1984, ApJ, 286, 644
\bibitem[Olive, Skillman \& Steigman 1997]{oss}
Olive, K.A. Skillman, E.D., \& Steigman, G. 1997, ApJ, 489, 1006
\bibitem[\"Ostlin 2000]{o00}
\"Ostlin, G. 2000, ApJ, 535, L99
\bibitem[\"Ostlin et al. 2001]{oab_p}
\"Ostlin, G., Amram, P., Bergvall, N., et al.  2001, A\&A, 374, 800
\bibitem[Petrosian et al. 1997]{p97}
Petrosian, A.R., Boulesteix, J., Comte, G., Kunth, D., \& LeCoarer E. 
1997, A\&A, 318, 390
\bibitem[Pettini \& Lipman 1995]{pl_p}
Pettini, M., \& Lipman, K. 1995, A\&A, 297, L63
\bibitem[Pilyugin 1993]{pil}
Pilyugin, L.S. 1993, A\&A 277, 42
\bibitem[Recchi, Matteucci \& D'Ercole 2001]{p1} 
Recchi, S., Matteucci, F., \& D'Ercole, A. MNRAS, 322, 800 (Paper I)
\bibitem[Reimers 1975]{r75}
Reimers, D. 1975, Mem. R. Sci. Liege Ser. 6, 8, 369
\bibitem[Renzini \& Voli 1981]{rv_p}
Renzini, A., \& Voli, M. 1981, A\&A, 94, 175 (RV81)
\bibitem[Romano et al. 2000]{donatella}
Romano, D., Matteucci, F., Salucci, P., \& Chiappini, C. 2000, ApJ, 539, 235
\bibitem[Sait\=o et al. 1992]{s92}
Sait\=o, M., Sasaki, M., Ohta, K., \& Yamada, T. 1992, PASJ, 44, 593
\bibitem[Salpeter 1955]{s_p}
Salpeter, E.E. 1955, ApJ, 121, 161
\bibitem[Searle \& Sargent 1972]{ss} 
Searle, L., \& Sargent, W.L.W. 1972, ApJ, 173, 25
\bibitem[Schulte-Ladbeck et al. 2001]{regina}
Schulte-Ladbeck, R.E., Hopp, U., Greggio, L., Crone, M.M., \& Drozdovsky,
        I.O. 2001, AJ, in press
\bibitem[Skillman \& Kennicutt 1993]{sk}
Skillman, E.D., \& Kennicutt, R.C.J. 1993, ApJ, 411, 655
\bibitem[Skillman, Terlevich \& Terlevich 1998]{stt}
Skillman, E.D., Terlevich, E., \& Terlevich R. 1998, in Primordial 
Nuclei and their Galactic Evolution, ed. N. Prantzos, M. Tosi 
\& R. van Steiger (Dordrecht: Kluwer), p. 105
\bibitem[Skillman et al. 1988]{s88}
Skillman, E.D., Terlevich, R., Teuben, P.J., \& van Woerden, H. 
1988, A\&A, 198, 33
\bibitem[Terlevich, Skillman \& Terlevich 1996]{tst}
Terlevich, E., Skillman, E.D., \& Terlevich, R. 1996, in The Interplay 
between Massive Star Formation, the ISM and Galaxy Evolution, eds. 
D. Kunth, B. Guiderdoni, M. Heydari-Malayeri and T. Thuan 
(Gif-Sur-Yvette: Ed. Fronti\`eres), p. 395
\bibitem[Timmes, Woosley \& Weaver 1995]{tww}
Timmes, F.X., Woosley, S.E., \& Weaver, T.A. 1995, ApJS, 98, 617
\bibitem[Tosi 1998]{t98}
Tosi, M. 1999, in Dwarf galaxies and cosmology, T.X.Thuan, C.Balkowski, 
 V.Cayatte, J.Tran Thanh Van eds 
 (Edition Fronti\`eres, Gif-sur-Yvette, France), p.443
\bibitem[Tosi 2001]{t01}
Tosi, M. 2001, in Dwarf Galaxies and their Environment, K.S. de Boer, R.J. 
  Dettmar, U. Klein eds (Shaker Verlag, De), in press
\bibitem[Tosi et al. 1991]{tgmf}
Tosi, M., Greggio, L., Marconi, G., \& Focardi, P. 1991, AJ 102, 951
\bibitem[van den Hoek \& Groenewegen 1997]{vg_p}
van den Hoek, L.B., \& Groenewegen, M.A.T. 1997, A\&AS, 123, 305 (VG97)
\bibitem[van Zee 2001]{v01}
van Zee, L. 2001, AJ, 121, 2003
\bibitem[van Zee, Westpfahl \& Haynes 1998]{vwh}
van Zee, L., Westpfahl, D., \& Haynes M.P. 1998, AJ, 115, 1000
\bibitem[Viallefond, Lequeux \& Comte 1987]{vlc}
Viallefond, F., Lequeux, J., \& Comte G., 1987, in Starbursts and 
Galaxy Evolution, Proc. of the Twenty-second Moriond Astrophysics 
Meeting (Gif-Sur-Yvette: Ed. Fronti\`eres), p. 139
\bibitem[Vidal-Madjar et al. 2000]{v00}
Vidal-Madjar, A., Kunth, D., Lecavelier des Etangs, A., et al. 
2000, ApJ, 538, L77
\bibitem[Vilchez \& Iglesias-P\'aramo 1998]{vi_p}
Vilchez, J.M., \& Iglesias-P\'aramo, J. 1998, ApJ, 508, 248
\bibitem[Woosley \& Weaver 1995]{ww_p}
Woosley, S.E., \& Weaver, T.A. 1995, ApJS, 101, 181 (WW) 

\end{thebibliography}
\end{document}